\documentclass{elsart}
\usepackage[numbers,sort&compress]{natbib}
\usepackage{color,setspace,times}
\usepackage{amssymb,natbib,amsmath,graphicx,color,rotating,subfigure,url}
\usepackage{lineno}

\journal{Physica A}

\begin{document}

\begin{frontmatter}

\title{Dynamics of Bid-ask Spread Return and Volatility of the Chinese Stock Market}
\author[SE]{Tian Qiu \corauthref{cor}},
\corauth[cor]{Corresponding author. Address: 696 South Fenghe
Avenue, School of Information Engineering, Nanchang Hangkong
University, Nanchang, 330063, China.} \ead{tianqiu.edu@gmail.com}
\author[SE]{Guang Chen},
\author[SA]{Li-Xin Zhong},
\author[SE]{Xiao-Run Wu}

\address[SE]{School of Information Engineering, Nanchang Hangkong University, Nanchang, 330063, China}
\address[SA]{School of Finance, Zhejiang University of Finance and Economics, Hangzhou, 310018, China}

\begin{abstract}

Bid-ask spread is taken as an important measure of the financial market liquidity. In this article, we study the dynamics of the spread return and the spread volatility of four liquid stocks in the Chinese stock market, including the memory effect and the multifractal nature. By investigating the autocorrelation function and the Detrended Fluctuation Analysis (DFA), we find that the spread return is lack of long-range memory, while the spread volatility is long-range time correlated. Moreover, by applying the Multifractal Detrended Fluctuation Analysis (MF-DFA), the spread return is observed to possess a strong multifractality, which is similar to the dynamics of a variety of financial quantities. Differently from the spread return, the spread volatility exhibits a weak multifractal nature.

\end{abstract}


\end{frontmatter}

\section{Introduction}


Financial dynamics has attracted a central interest of physicists in recent years \cite{man95,lux99,gia01,lal99,tot06,ren08,qiu10,mis10}. The financial market has accumulated a large amount of historical data in the past years, which provides an abundant data resource for the empirical study. Some scaling behaviors have been revealed to be universal for different markets \cite{man95,lux99,gia01}, e.g., the well known ''inverse cubic law'' of the probability distribution of the price return. The so-called ''volatility clustering'' is also found for a variety of financial markets  \cite{liu99,gop00}, which refers to the long-range time correlation of the absolute value of the price return. Differently from those universal behaviors, unique characteristics are found for several emerging markets. A typical example is the ''anti-leverage effect'' of the Chinese stock market \cite{qiu06}, which indicates an anti-effect from the ''leverage effect'' of most mature financial markets in the return-volatility correlation \cite {bou01}. On the other hand, the financial markets can be taken as a complex system composed of many elements, with the scaling behavior emerging from the many-body interactions. Different models and theoretical approaches have been developed to describe financial markets \cite{cha97,sta99,con00,egu00,wan09}.

Within this framework, the so-called bid-ask spread is served as an important indicator to quantify the financial market liquidity and efficiency. Most modern financial markets are order-driven markets, which adopt the continuous double auction mechanism. There are two basic types of orders, the market order and the limit order. For the market order, trade is carried out immediately when receiving the order. For the limit order, order is stored, and it waits for a best price for trading. The buy limit order is called bid, and the sell limit order is called ask. The bid-ask spread refers to the price difference between the lowest ask price and the highest bid price. Generally speaking, the lower the spread, the smaller the transaction cost, and also a higher liquidity of the market.

So far, statistical properties of the bid-ask spread of the limit-orders have been widely studied \cite{ple05,mik08,far04,gu07a}. Empirical study shows that, the probability distribution of the bid-ask spread obeys a power law behavior, with the exponent around $3.0$ \cite {mik08, far04}. Long-range time correlation of the bid-ask spread is also revealed for different markets, including the Chinese stock market \cite {mik08,gu07a}. In addition, differently from the multifractality of a variety of financial quantities, the bid-ask spread is reported to be mono-fractal for the Chinese stock market \cite{gu07a}. On the other hand, a lot of work contributes to the order flow dynamics, and the microscopic mechanism of price formation    \cite{bou02,pot03,dan03,far04,lil05,ple05,lil07,mik08,tot09}. However, the dynamics of the spread return and the spread volatility has not been analyzed or reported in detail, to our knowledge. The spread return is defined as the difference of bid-ask spread between two successive time point, and the spread volatility is defined as the magnitude of the spread return. In fact, in finance, one concerns more on the price return or the price volatility, rather than the price itself. Similarly, it is essential to investigate the bid-ask spread return and the spread volatility dynamics.

With this incentive, in this paper, we study the dynamics of the spread return and the spread volatility. Based on the tick-by-tick data of four liquid stocks in the Chinese stock market, we investigate the memory effect of the spread return and the spread volatility by applying the autocorrelation function and the Detrended Fluctuation Analysis (DFA). Multifractal nature of the spread return and the spread volatility is also revealed, by employing the Multifractal Detrended Fluctuation Analysis (MF-DFA).

The remainder of this paper is organized as follows. In the next section, we present the datasets we analyzed. In section 3, we introduce the definition of the bid-ask spread, the spread return, and the spread volatility, and then we focus on the memory effect of the spread return and the spread volatility by applying the autocorrelation function and the DFA method in section 4. In section 5, we emphasize the multifractal nature of the spread return and the spread volatility. Finally comes the conclusion.

\section{Datasets}

The Chinese stock market is an order-driven market based on the continuous double auction \cite {zho04,gu07a,gu08a}. Before July 1, 2006, only limit orders were permitted to submit in the order placement. The Chinese stock market is composed of two stock exchanges, the Shanghai Stock Exchange (SHSE) and the Shenzhen Stock Exchange (SZSE). The SHSE was established on November 26, 1990, and the SZSE was established on December 1, 1990. The dataset we analyzed is based on the tick-by-tick data of 4 liquid stocks listed on the SHSE and the SZSE, i.e., the Shenzhen Development Bank Co., Ltd (SDB) stock, the Guangdong Electric Power Development Co., Ltd. (GEP) stock, the Datang Telecom Co., Ltd (DTT) stock, and the Shanghai Lujiazui Finance and Trade Zone Co., Ltd (SLFT) stock. The time scale of the datasets covers 3 whole year from 2004 to 2006, and the time resolution of the data record is about 15 seconds on average.

\section{Definition of the Spread Return and the Spread Volatility}

Different definition of the bid-ask spread has been proposed \cite{rol84,sto89,hua97,dan03,far04,ple05,far05,caj07,wya06}. In our study, we define the bid-ask spread $\widetilde{S}$ as the price difference between the lowest ask price and the highest bid price, which reads,

\begin{equation}
\widetilde{S}(t')=a(t')-b(t'),
\end{equation}

where $a(t')$ and $b(t')$ are the lowest ask price and the highest bid price at time $t'$, respectively. To quantify the spread $\widetilde{S}$ at a same time scale, we define a rescaled spread $S$ in a time interval $\Delta t$ as \cite{ple05},

\begin{equation}
S(t')=\frac{1}{N}\sum_{i=1}^{N}\widetilde{S}_{i}(t'),
\end{equation}

where $N$ is the total number of transactions in the time interval $\Delta t$. In this article, we study the time scale $\Delta t=1$ minute. The original spread return is defined as the spread change between two consecutive values of the rescaled spread,

\begin{equation}
\widetilde{R}(t')=ln (\frac{S(t')}{S(t'-1)}),
\end{equation}

The original spread volatility is then defined as the absolute value of the original spread return, i.e., $\widetilde{V}(t')=|\widetilde{R}(t')|$.

It has been reported that the bid-ask spread shows similar characteristics as a number of financial quantities, such as the long-range time-correlation of the spread series. However, the spread is also found to present some unique feature. For example, the spread exhibits a mono-fractal characteristic \cite{gu07a}, which is different from the multifractality of most financial quantities. What feature could the spread return and the spread volatility present? In the following section, we investigate the memory effect of the spread return and the spread volatility by the autocorrelation function and the DFA method.

\begin{figure}[htb]
\centering
\includegraphics[width=8.5cm]{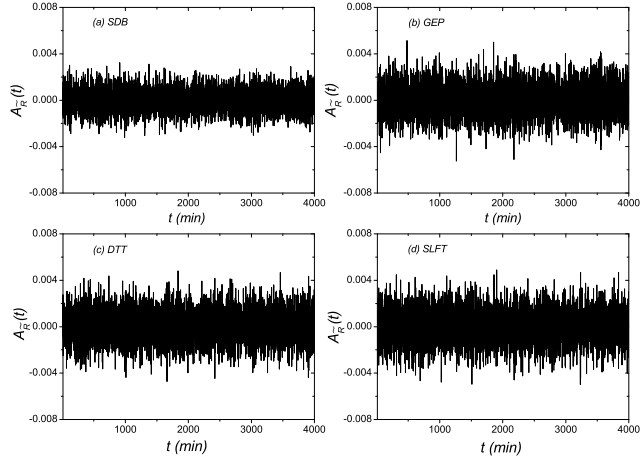}
\caption{\label{Fig:1} The autocorrelation function $A_{\widetilde{R}}(t)$ of the original spread return $\widetilde{R}(t')$ is displayed.}
\end{figure}

\section{Memory Effect}

\subsection{Autocorrelation}

It is well known that the price return exhibits a short-range time-correlation, while the price volatility is long-range correlated. To detect the memory effect of the spread return and the spread volatility, we investigate the autocorrelation function, which is defined as,

\begin{equation}
A_{\widetilde{R}}(t)=[\langle \widetilde{R}(t')\widetilde{R}(t+t') \rangle-\langle \widetilde{R}(t') \rangle ^{2}]/\sigma_{\widetilde{R}}^{2}, \label{3}
\end{equation}

\begin{equation}
A_{\widetilde{V}}(t)=[\langle \widetilde{V}(t')\widetilde{V}(t+t') \rangle-\langle \widetilde{V}(t') \rangle ^{2}]/\sigma_{\widetilde{V}}^{2}, \label{3}
\end{equation}

where $\sigma_{\widetilde{R}}^{2}=\langle \widetilde{R}(t')^{2} \rangle-\langle \widetilde{R}(t')
\rangle^{2}$, and $\sigma_{\widetilde{V}}^{2}=\langle \widetilde{V}(t')^{2} \rangle-\langle \widetilde{V}(t')
\rangle^{2}$, and the $\langle \cdot\cdot\cdot \rangle$ takes an average over $t'$.

\begin{figure}[htb]
\centering
\includegraphics[width=8.5cm]{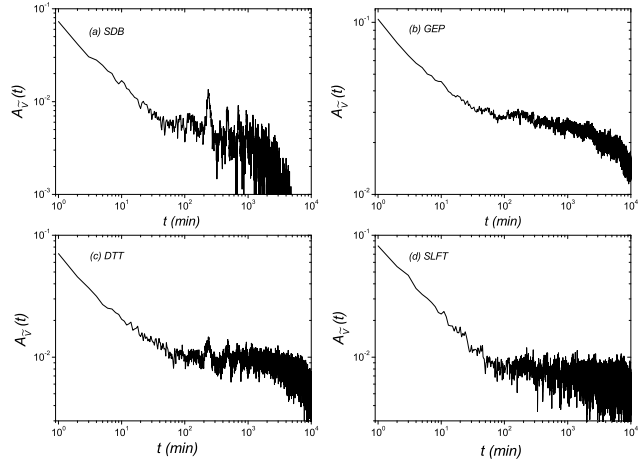}
\caption{\label{Fig:1}  The autocorrelation function $A_{\widetilde{V}}(t)$ of the original spread volatility $\widetilde{V}(t')$ is displayed on a log-log scale.}
\end{figure}

In Fig. 1, the autocorrelation function of the original spread return $\widetilde{R}(t')$ is displayed for the SDB, the GEP, the DTT, and the SLFT stock, respectively. It is observed that, for the four stocks, the autocorrelation of the spread return fluctuates around 0, which indicates a short-range time-correlation.

Fig. 2 shows the autocorrelation function of the original spread volatility $\widetilde{V}(t')$, where a periodic structure is observed. Such kind of periodic structure refers to the so-called intraday pattern, which has been found in a variety of high-frequent financial series \cite{dac93,woo85,har86,adm88,cha95,qiu09}, including the bid-ask spread \cite{gu07a,ni09,mci92}. In the Chinese stock market, there is an active opening call auction time period from 9:15 to 9:25 a.m., when the buy and sell orders are aggregated to match. The intraday pattern could be dominated by the active call auction at the opening time. The intraday pattern strongly influences the financial dynamics, therefore it requires an efficient method to remove the intraday pattern. A $\vartheta$-time scale method has been applied to remove the intraday-pattern of a FX market \cite{dac93}. In our study, we follow the method introduced by Ref. \cite{liu99}.

In the Chinese stock market, there are 240 consecutive working minutes in a trading day. We then segment the dataset of each trading day into 240 consecutive 1-min intervals. We define the intraday pattern of the spread return $M_{\widetilde{R}}$ and the spread volatility $M_{\widetilde{V}}$ at time $t'$ in these 240 working minutes as,

\begin{equation}
M_{\widetilde{R}}(t')=\frac{1}{N_{d}}\sum_{i=1}^{N_{d}}\widetilde{R}_{i}(t'),
\end{equation}

\begin{equation}
M_{\widetilde{V}}(t')=\frac{1}{N_{d}}\sum_{i=1}^{N_{d}}\widetilde{V}_{i}(t'),
\end{equation}

where $i$ is the $i^{th}$ trading day, $N_{d}$ is the number of trading days. The adjusted spread return $R$ and the adjusted spread volatility $V$ is then defined as,

\begin{equation}
R(t')=\frac{\widetilde{R}(t')}{M_{\widetilde{R}}(t')},
\end{equation}

\begin{equation}
V(t')=\frac{\widetilde{V}(t')}{M_{\widetilde{V}}(t')},
\end{equation}

The autocorrelation function $A_{V}(t)$ of the adjusted spread volatility is defined in a similar way as that of the original spread volatility, by replacing the original spread volatility $\widetilde{V}$ with the adjusted volatility $V$. After removing the intraday pattern, the autocorrelation function of the adjusted spread volatility is shown in Fig. 3, where most of the periodic peaks are eliminated. The $A_{V}(t)$ is found to decay slowly, which suggests a long-range time correlation of the spread volatility. However, the autocorrelation dynamics shows a large fluctuation, therefore it is difficult to detect the decay exponent of the $A_{V}(t)$.

\begin{figure}[htb]
\centering
\includegraphics[width=8.5cm]{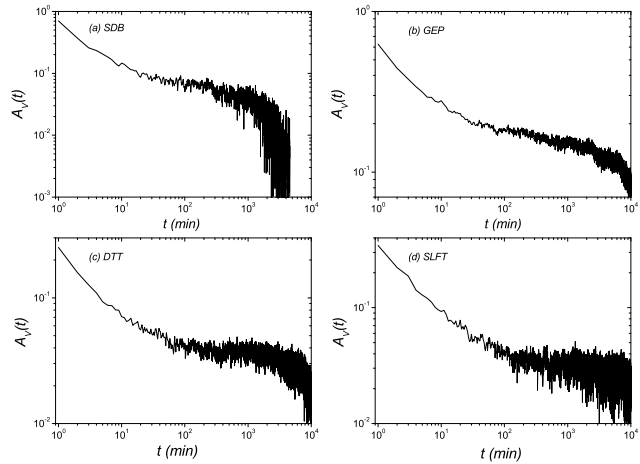}
\caption{\label{Fig:1}  The autocorrelation function $A_{V}(t)$ of the adjusted spread volatility $V(t')$ is displayed on a log-log scale.}
\end{figure}

\subsection{DFA Method}

For the non-stationary time series, a widely adopted measurement of the time-correlation is the DFA method \cite {pen95, pen94}, which has been successfully applied to detect the time correlations in various physical systems. We then introduce the DFA method.

Considering a fluctuating time series $A(t')$, one can construct a new time series by,
\begin{equation}
B(t')=\sum\limits_{t''=1}^{t'} [A(t'')-A_{ave}], \label{e8}
\end{equation}

where $A_{ave}$ is the average of the $A(t')$ in the total time interval $[1,T]$. By dividing the total time interval into windows $N_t$ with a size of $t$, and linearly fit $B(t')$ to a linear function $B_{t}(t')$ in
each window, the DFA function of the $k_{th}$ window box is defined by,

\begin{equation}
f_{k}(t)^{2}=\frac{1}{t}\sum\limits_{t'=(k-1)t+1}^{kt}{[B(t')-B_t(t')]}^2
, \label{e9}
\end{equation}

The overall detrended fluctuation is then estimated as,

\begin{equation}
F_{2}(t)^{2}=\frac{1}{N_t}\sum\limits_{k=1}^{N_t}{[f_k(t)]}^2 ,
\label{e9}
\end{equation}

In general, $ F_{2}(t)$ scales by a power-law  $F_{2}(t)\sim t^{H}$. If $0.5<H<1.0$, $A(t')$ is considered to be long-range correlated in time; if $0<H<0.5$, $A(t')$ is temporally anti-correlated; $H=0.5$ corresponds to the Gaussian white noise, while $H=1.0$ suggests the $1/f$ noise of the $A(t')$ series. If $H$ is bigger than $1.0$, the $A(t')$ series is taken as an unstable series.

\begin{figure}[htb]
\centering
\includegraphics[width=8.5cm]{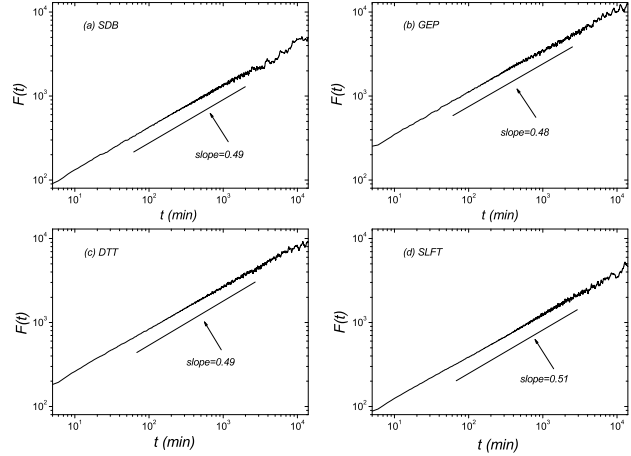}
\caption{\label{Fig:1} The DFA function $F(t)$ of the adjusted spread return $R$ is displayed on a log-log scale.}
\end{figure}

We apply the DFA to the adjusted spread return and the adjusted spread volatility. In Fig. 4, the DFA function of the adjusted spread return is shown on a log-log scale. A power-law increase is observed, with the exponent is measured to 0.49 for the SDB, 0.48 for the GEP, 0.49 for the DTT, and 0.51 for the SLFT. The DFA exponents of the four stocks are all around $0.50$, which indicates a Gaussian white noise of the adjusted spread return, i.e., no memory in the spread return series. The results of the DFA method are consistent with those obtained from the autocorrelation functions of the spread return. The absence of memory in the spread return series is similar as that of the price return. Differently from the spread return, the DFA of the adjusted spread volatility presents a cross-over phenomenon for all the four stocks, as shown in Fig. 5. For the forepart time scale, the DFA exponents are measured to be 0.68 for the SDB, 0.73 for the GEP, 0.65 for the DTT, and 0.66 for the SLFT. The value of the four exponents is ranged inbetween $(0.5,1)$, which indicates a long-range correlation of the spread volatility. For the afterward time scale, DFA exponents are measured to be 1.00 for the SDB, 1.07 for the GEP, 1.11 for the DTT, and 0.81 for the SLFT. Except for the stock SLFT, the other 3 stocks present a $1/f$ noise characteristic. Such kind of cross-over behavior of the spread volatility is also similar to the price volatility dynamics.

To this end, the spread return is found to be lack of long-range memory, whereas the spread volatility exhibits a long-range time correlation. The memory effect of the spread return and the spread volatility is quite similar to that of the price return and the price volatility, respectively.

\begin{figure}[htb]
\centering
\includegraphics[width=8.5cm]{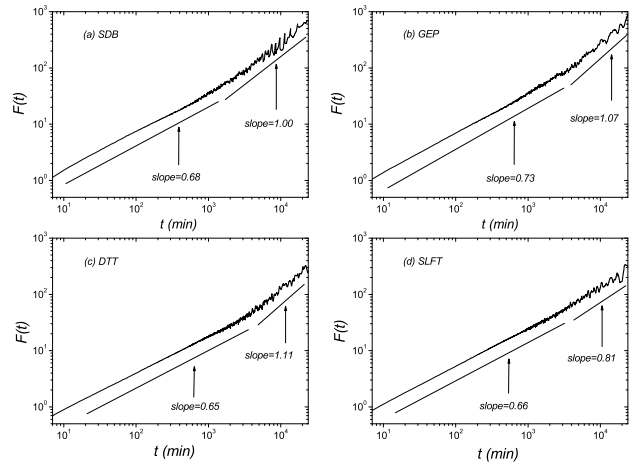}
\caption{\label{Fig:1}  The DFA function $F(t)$ of the adjusted spread volatility $V$ is displayed on a log-log scale.}
\end{figure}

\section{Multifractal Analysis}

The multifractal analysis is widely used to evaluate the risk of the financial market \cite{wei08}, also to quantify the market inefficiency \cite{zun08}. Multifractal characteristic has been found in many quantities of financial markets, such as the price volatility, and the trading volume \cite{jia08a,zho09,moy06}. However, differently from most of the financial quantities, the bid-ask spread has been reported to be mono-fractal. Could the spread return and the spread volatility possess the multifractal nature? In our study, we apply the MF-DFA to reveal the multifractal feature of the spread return and the spread volatility. The MF-DFA is a generalization of the original DFA method. It considers different orders of the detrended fluctuations, which reads,

\begin{equation}
F_{q}(t) =\{ \frac{1 }{N_t}\sum_{k=1}^{N_t}[f_{k}(t)]^q\}^{1/q},
\label{e400}
\end{equation}

where $q$ is the $q_{th}$ order of the detrended fluctuation. If $q=2$, the MF-DFA then reduces to the original DFA method. The $q$ can take any real number except $q = 0$, where $F_{0}(t)$ reads by,

\begin{equation}
F_{0}(t) =\exp\{ \frac{1 }{N_t}\sum_{k=1}^{N_t}\ln[f_{k}(t)]\},
\label{e500}
\end{equation}

Generally speaking, the MF-DFA function $F_q(t)$ scales with the window size $t$ by,

\begin{equation}
F_{q}(t)\sim t^{h(q)}, \label{e600}
\end{equation}

where $h(q)$ is the MF-DFA exponent. For a large absolute value of $q$, the $F_{q}(t)$ shows a large fluctuation with $t$, due to the finite size of the investigated financial system. Therefore, in our study, we investigate the multifractality of the spread return and the spread volatility with $q \in [-6, 6]$.

\begin{figure}[htb]
\centering
\includegraphics[width=8.5cm]{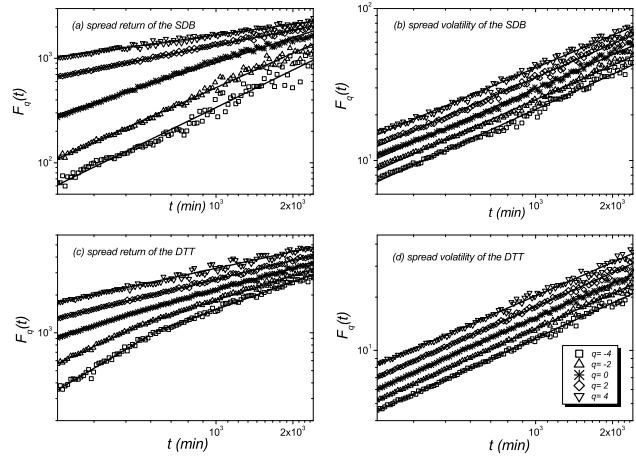}
\caption{\label{Fig:1} (a) is the MF-DFA function $F_q(t)$ of the adjusted spread return for the SDB stock, with the exponents measured to be 1.20, 1.10, 0.80, 0.49, and 0.35 for $q=$ -4, -2, 0, 2, and 4, respectively. (b) is the $F_q(t)$ of the adjusted spread volatility for the SDB stock, with the exponents measured to be 0.77, 0.75, 0.74, 0.73, and 0.71 for $q=$ -4, -2, 0, 2, and 4, respectively. (c) is the MF-DFA function $F_q(t)$ of the adjusted spread return for the DTT stock, with the exponents measured to be 0.59, 0.50, and 0.44 for $q=$ 0, 2, and 4, respectively. For $q=-4$ and $-2$, it shows a cross-over behavior, with exponent measured to be 1.21 and 0.96 for the forepart time scale, and 0.74 and 0.65 for the afterward time scale, respectively. (d) is the $F_q(t)$ of the adjusted spread volatility for the DTT stock, with the exponents measured to be 0.66, 0.66, 0.65, 0.64, and 0.63 for $q=$ -4, -2, 0, 2, and 4, respectively. }
\end{figure}

In Fig. 6, the $F_q(t)$ of the adjusted spread return $R$ and the adjusted spread volatility $V$ is shown for the SDB stock and the DTT stock. It is observed that, the $F_q(t)$ of both the adjusted spread return $R$ and the adjusted spread volatility $V$ shows a power law scaling for the investigated value of $q$, except for a two-stage power law scaling in the $F_q(t)$ of the adjusted spread return $R$ of the DTT stock at $q=-2,-4$. The cross-over behavior of the DTT stock might be from fluctuations induced by the big value of $q$. From Fig. 6 (a), the exponent of the $F_q(t)$ is estimated to be 1.20, 1.10, 0.80, 0.49, and 0.35 for the $q=$ -4, -2, 0, 2, and 4 for the SDB stock, respectively. That is to say, the increasing trend of the $F_q(t)$ with $t$ becomes more slowly as $q$ increases. Similar behavior is also found for the DTT stock. It indicates a multifractal nature of the spread return. For the adjusted spread volatility, the exponent of the $F_q(t)$ is estimated to be 0.77, 0.75, 0.74, 0.73, and 0.71 for $q=$ -4, -2, 0, 2, and 4 for the SDB stock, respectively. The exponent of the $F_q(t)$ decreases very slowly with $q$. Also, similar behavior is observed for the DTT stock. It suggests a weak multifractal feature of the spread volatility.

To further understand the multifractal nature of the spread return and the spread volatility, we investigate the MF-DFA exponent $h(q)$ on $q$, the scaling exponent function $\tau(q)$ on $q$, and the multifractal singularity spectrum $f(\alpha)$ on $\alpha$, for the adjusted spread return and the adjusted spread volatility. We take the SDB stock as an example to illustrate the results. The MF-DFA exponent $h(q)$ on $q$ of the the adjusted spread return $R$ and the adjusted spread volatility $V$ of the stock SDB is shown in Fig.7 (a) and (d). For a mono-fractal series, one has $\Delta h(q)=h(q_{min})-h(q_{max})=0$. The $\Delta h=h(q_{-6})-h(q_{6})$ of the spread return and the spread volatility is estimated to be 1.02, and 0.15, respectively. It indicates that, the spread return shows a strong multifractality, whereas the spread volatility presents a weak multifractality. We then investigate the scaling exponent function $\tau(q)$ on $q$ based on partition function, which is also widely adopted to reveal the multifractality of a time series,

\begin{equation}
\tau(q)=qh(q)-D_{f}, \label{e700}
\end{equation}

where $D_f$ is the fractal dimension, with $D_f = 1$ in our case. For the mono-fractal, the $\tau(q)$ linearly increases with $q$. As shown in Fig. 7 (b), the $\tau(q)$ of the adjusted spread return presents a strong nonlinearity, which further suggests the multifractal nature of the spread return. For the adjusted spread volatility, as shown in Fig. 7 (e), the $\tau(q)$ nearly linearly increases with $q$, which again indicates a weak multifractality of the spread volatility. By the Legendre transformation, the local singularity exponent $\alpha$ and its spectrum $f(\alpha)$ can be calculated as \cite{hal86},

\begin{equation}
\alpha=d\tau(q)/dq, \label{e800}
\end{equation}
\begin{equation}
f(\alpha)=q\alpha-\tau(q), \label{e900}
\end{equation}

\begin{figure}[htb]
\centering
\includegraphics[width=8.5cm]{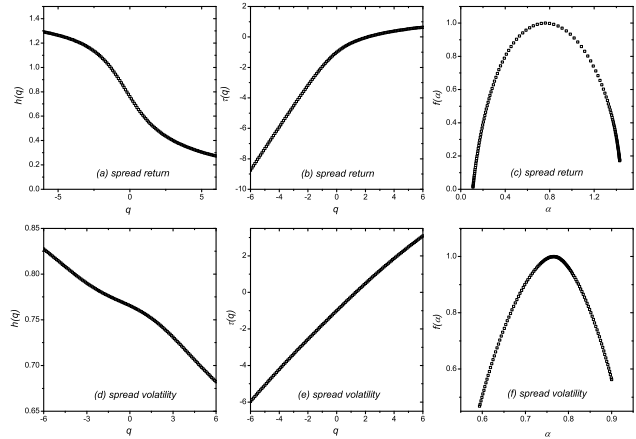}
\caption{\label{Fig:1} The MF-DFA exponent $h(q)$ on $q$ is shown in (a) and (d), and the scaling exponent function $\tau(q)$ on $q$ is shown in (b) and (e), and the multifractal singularity spectrum $f(\alpha)$ on $\alpha$ is shown in (c) and (f). The upper panels are for the adjusted spread return, and the lower panels are for the adjusted spread volatility of the SDB stock, respectively. }
\end{figure}

The difference between the maximum and the minimum of the local singularity exponent $\Delta\alpha \triangleq \alpha_{max}-\alpha_{min}$ indicates the width of the extracted multifractal spectrum. The larger the $\Delta\alpha$, the stronger the multifractality. Fig. 7 (c) and (f) illustrate the multifractal singularity spectra $f(\alpha)$ of the adjusted spread return $R$ and the adjusted spread volatility $V$. The width of the extracted multifractal spectrum $\Delta\alpha$ is measured to be $1.32$ for the spread return $R$, indicating a strong multifractality, and $0.31$ for the spread volatility $V$, indicating a weak multifractality. The results of $f(\alpha)$ further support the results obtained from the MF-DFA exponents $h(q)$ and the scaling exponent function $\tau(q)$, i.e., the spread return shows a strong multifractality, whereas the spread volatility presents a weak multifractality.

\section{Conclusion}

In finance, one of the central tasks is to evaluate the market efficiency. The bid-ask spread is served as an important indictor to quantify the financial market liquidity and vitality. In the present article, based on the tick-by-tick data of four liquid stocks in the Chinese stock market, we investigate the memory effect and the multifractal nature of the spread return and the spread volatility. It is observed that, the spread volatility shows a significant long-range time-correlation, while the spread return is short-range time-correlated. Moreover, the spread return is found to exhibit a strong multifractal characteristic, which has been observed in dynamics of a number of financial quantities. Differently from the spread return, the spread volatility presents a weak multifractal feature. Our work might help to better understand the financial market.

\bigskip
{\textbf{Acknowledgments:}}

This work was partially supported by the National Natural Science Foundation of China under Grant Nos. 10805025 and 11175079, the General Program of Social Science Research
Fund of Ministry of Education of China under Grant No. 10YJAZH137, the Natural Science Foundation of Zhejiang Province under Grant No. Y6110687.

\bibliography{rint}
\end{document}